\documentclass[twocolumn,showpacs,preprintnumbers,amsmath,amssymb]{revtex4}
\usepackage{graphicx}

\begin{document}

\title{Tkachenko waves in rapidly rotating Bose-Einstein condensates}

\author{L. O. Baksmaty, S. J. Woo, S. Choi, N. P. Bigelow}
\affiliation{Department of Physics and Astronomy, and Laboratory for  Laser
Energetics, University of Rochester, Rochester, NY 14627}
\date{\today}

\begin{abstract}
We present a mean-field theory numerical study of Tkachenko waves of a vortex lattice
in trapped atomic Bose-Einstein condenstates. 
Our results show remarkable qualitative and quantitative agreement with
recent experiments at JILA. We extend our calculations beyond the conditions of the experiment, probing
deeper into the incompressible regime where we find excellent agreement with analytical results. 
In addition, bulk excitations observed in the experiment are discussed.

\end{abstract}
 \maketitle

The dynamics of quantized vorticies are important in a wide variety of physical phenomena from turbulence to neutron stars, to superconductivity and superfluidity~\cite{sonin}. 
The broad relevance of this problem is partly responsible
for the current interest in vortex lattices in trapped Bose-Einstein condensates
(BECs). This system is also attractive because it can easily be manipulated and imaged~\cite{mit,jilav,jilat} and because the condensates is accurately described within 
mean-field theory. Until recently much of the progress made in the study of vortex arrays in neutral superfluids 
was in the context of $^4$He. This system however is notoriously difficult to manipulate and control. 
Vortex lattices in BECs have thus created an unprecendented opportunity to study vortex matter
in detail. This has been recently demonstrated in spectacular experiments at JILA in which
Tkachenko waves as well as hydrodynamic shape oscillations were directly observed~\cite{jilat}.  In this paper we present a numerical study of these vortex matter excitations.  Our approach has the advantage that it does not require any assumptions about the compressibility or homogeneity of the condensate.  
   
In the limiting case of an incompressible irrotational fluid, Tkachenko waves are 
transverse vortex-displacement waves traveling in the triangular lattice of vortex lines which constitute the 
ground state of the rotating superfluid~\cite{tkachenko}. 
In this idealized situation, the size of the vortex core is infinitesimal and the 
dynamics of the fluid density may be ignored.
 For a simple illustration, consider a system of $n$ vortices each of circulation $k$, in an irrotational, 
incompressible fluid of density $\rho$ contained by a vessel rotating at angular frequency $\Omega$.
Let $\bf v\left(\bf r\right)$ represent the fluid velocity at position $\bf r$. 
If the position of the $i$th vortex is labeled by a complex number $z_{i}$, the free energy 
($f$) of the array may be written as 
\begin{equation}
f=\frac{\rho k^2}{4\pi}\left[-\frac{1}{2}\sum_{i=1}^{n}\sum_{j=1}^{n}\log{\mid z_{i}-z_{j} 
\mid}^2-\Omega \sum_{i=1}^{n}(1-\mid z_{i} \mid^2)\right]. 
\label{free}
\end{equation} 
Here we ignore the effects of the boundaries and the energy associated with 
vortex cores~\cite{sonin,ziff,campbell}. 
By taking into account the irrotational ($\nabla\times {\bf v}=0$) and imcompressible
($\nabla \cdot {\bf v}= 0$) nature of the fluid,  
we may derive this expression from the simple classical relation for the energy which in 
this case is purely kinetic:   
 \begin{equation}
f=\rho\int \left[\frac{1}{2}{\bf v\left(r\right)}^2-\Omega
\left({\bf r}\times{\bf v\left(r\right)}\right)\right]d{\bf r}.  
\label{classical}
\end{equation}
We can observe from Eq.~(\ref{free}) that the vortex lattice, which represents 
a local minimum of this free energy, results from a competition between two terms: 
the logarithmic term which represents the intervortex repulsion, and the quadratic rotational term 
which pulls the vortices towards the 
center. The Tkachenko waves are organized oscillations of the vortex positions about this equilibrium. 
On the microscopic level, the vortices precess about their equilibrium position 
in elliptical orbits against the trap rotation with the major axis perpendicular to the wave propagation. 
On the macroscopic scale, the array is 
seen to undergo harmonic distortions which shear the lattice and cause the rotation of the array to alternatively 
slow down and speed up. 
    
Vortex arrays produced in rapidly rotating trapped BECs depart from this idealized case
in two important aspects: the system is significantly compressible, and the underlying fluid is 
inhomogeneous. Due to finite compressibility, it is important to consider the 
dynamics of the underlying fluid for an accurate treatment. This is especially true in the 
light of recent interest in lattices in the mean-field quantum Hall 
regime~\cite{ho,eric}. To date, most vortex lattice studies rely on incompressible hydrodynamic 
approximations~\cite{sonin,ziff,campbell,anglin}. The hydrodynamic approximation restricts 
the calculation to longwavelength excitations and the assumption of incompressibility ignores
the dynamics of the fluid density. Baym~\cite{baym} has recently demonstrated the importance
of compressiblity to the description of current experimental data. 
The compressibility of the lattice may be gauged from the ratio $\xi^2 / \epsilon^2$. 
Here $\xi$ and $\epsilon$ are
the healing length and intervortex distance respectively. The healing length $\xi$ estimates the size of the vortex core.  We define these  
as $\xi=1/\sqrt{8\pi na}$ and $\epsilon^2=2\pi / (\sqrt{3} \Omega)$~\cite{svid}~, where 
$a$ is the scattering length of the trapped species, $n$ is the density at the center  of the condensate and $\Omega$ is the angular frequency of the of the trap rotation.  In this Letter we bypass these familiar approximations by adopting 
a full numerical approach based on mean-field theory.   

As an initial step,  we must construct the undisturbed vortex array to a high degree of accuracy. We focus on the excitations of triangular vortex arrays with 
hexagonal symmetry, such as those obtained by recent experiments. 
A pancake shaped trap geometry is employed. We note that owing to the centrifugal reduction  in the radial trap frequency $\omega_{r}$, this is a reasonable assumption at
high rotation frequencies where most of our studies take place. 
More importantly, the pancake geometry has been experimentally justified by the 
high quality of direct imaging of high~\cite{jilab} and low energy lattice excitations -
if the curvature of vortex lines were significant, such imaging would not be possible. According 
to the $T=0$ mean field theory, the condensate
wavefunction in a frame rotating with angular velocity $\Omega$ satisfies the 
equation
\begin{equation}
i \hbar \frac{\partial \psi({\bf r}, t)}{\partial t} = \left[
-\frac{\hbar^2}{2m} \nabla^2 + V({\bf r}) +N g|\psi|^2
-L_{z}\Omega
\right] \psi({\bf r},t), 
\label{tdgpe}
\end{equation}
where $m$ is the mass of the species, $\omega_{r}$ is the radial
 trapping frequency, $N$ is the number of trapped atoms and 
$L_{z}$ is the angular momentum operator defined
 as $-i \hbar \frac{\partial }{\partial \phi}$. 
The coupling constant $g$ and the external 
potential V({\bf r}) are defined as $g=4\pi a \hbar^2/m$ 
and $V({\bf r})=\frac{1}{2}m \omega_{r}^2 r^2$ respectively. 
The corresponding stationary equation is solved using a steepest descent  
technique starting from a triangular seed array. We define the fluctuation of the order parameter $\delta\psi$ by, 
\begin{equation}
\delta\psi=u({\bf r})e^{-i\omega t}-v({\bf r})^{*} e^{i\omega t}.
\label{ansatz}
\end{equation}
Following the Bogoliubov-de Gennes procedure,  first order expansion around the ground state in terms of   
$\delta \psi$ then yields the eigen-equation for the exciation amplitudes 
$u(\bf r)$ and $v(\bf r)$~\cite{svid}:    
\begin{widetext}
\begin{equation}
\left(
\begin{array}{cc}
-\frac{\hbar^2}{2m} \nabla^2 + V({\bf r})+2N g|\psi|^2-Lz\Omega-\mu
 & gN\psi^2 \\
gN\psi^{*2}&-\frac{\hbar^2}{2m} \nabla^2 + V({\bf r})+2N g|\psi|^2+Lz\Omega-\mu
\end{array}
\right)\left( \begin{array}{c}u\\v \end{array}\right)=
\omega\left( \begin{array}{c}u\\-v \end{array}\right). 
\label{eval}
\end{equation}
\end{widetext}
Unless otherwise stated, we express energy, length and time in the oscillator
units $\hbar\omega_{r}$, $\sqrt{\hbar /m\omega_{r}}$
and $1/\omega_{r}$ respectively. 
For the solution of the coupled eigenvalue equation(Eq. (\ref{eval})), 
we use a finite element method~\cite{fem}. The 
eigenvalue problem is characterized by three length scales: the Thomas-Fermi
radius $R_{TF}=\sqrt{2\mu}$, the healing length $\xi$, and the 
inter-vortex spacing $\epsilon$. 
$R_{TF}$ specifies the extent of the lattice, and $\xi$ and $\epsilon$ constrain
 the density of the
grid required by the calculation. Care must be taken to ensure the grid  
captures features over these length scales and this renders the problem very 
computationally expensive. It was not unusual to obtain sparse matrices of the order
$10^{5} {\times} 10^{5}$; to solve such large eigensystems, we used a
variety of numerical libraries and visualization tools~\cite{numerics}. 
 
Following Ref.~\cite{anglin}, 
we represent a Tkachenko wave by a complex valued function 
$D_{n,m}\left({\bf r}, t\right)$ defined by 
\begin{eqnarray}
D_{n,m}({\bf r}, t)=e^{i\phi}[f_{n,m}^{a}(r)\sin{(m\phi-\omega t-\alpha)}
\\+if_{n,m}^{b}(r)\cos{(m\phi-\omega t-\alpha)}] \nonumber
\label{d}
\end{eqnarray}
The integer pair $(n,m)$ represent the radial and angular order of the excitations, 
${\bf r}=(r,\phi)$, $\alpha$ is an arbitrary phase and $\{(f_{n,m}^{a}(r),f_{n,m}^{b}(r))\}$ 
are real valued functions. The real and imaginary part of $D_{n,m}$, represent 
the displacement of a vortex at $\bf r$ along the $x$ and $y$ axis respectively. 
According to $D_{n,m}$, every vortex moves in an elliptical path. 

\begin{figure}
\includegraphics[width=8cm,height=4cm]{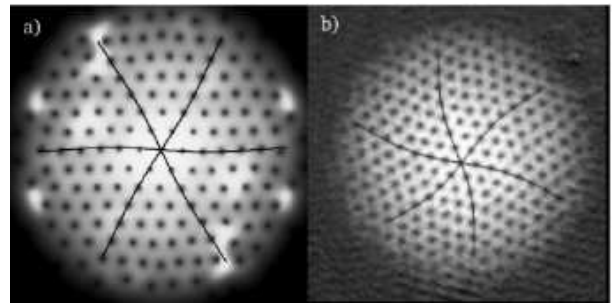}
\caption{Both figures depict a snapshot of the (1,0) mode in action. 
a) Simulation. b) Result from the JILA experiment. The curve is a fitted sine 
wave of wavelength 1.33$R_{TF}$}
\label{jila_rochester_10}
\end{figure}

\begin{figure}
\includegraphics[width=8cm,height=4cm]{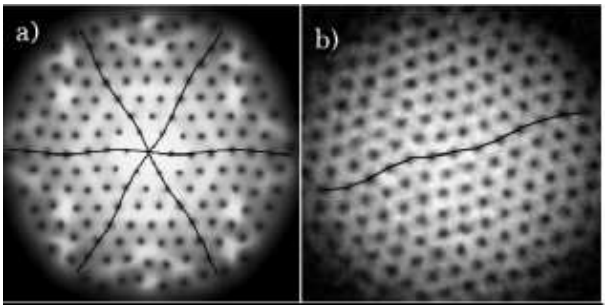}
\caption{(2,0) mode in action. a) Simulation. b) Result from the JILA experiment. 
In both cases, the curves 
are drawn to guide the eye.}
\label{jila_rochester_20}
\end{figure}

One of the challenges 
of comparing our numerical calculation to experiment or to analytic results 
is the proper identification of the excitations obtained. In our calculation, the degree of freedom is the fluctuation of the field amplitude $\delta\psi$, while in analytical work it is the distortion $D_{n,m}$ of the vortex array that is
employed. One approach, 
involves making ``movies" of each excitation by 
using Eq.~(\ref{ansatz}). In Figs.~\ref{jila_rochester_10} and \ref{jila_rochester_20},
 we show snapshots from 
two such movies for the $(1,0)$ and $(2, 0)$ modes and compare them to experimental data. 
 \begin{figure}
\includegraphics[width=9.0cm,height=5.0cm]{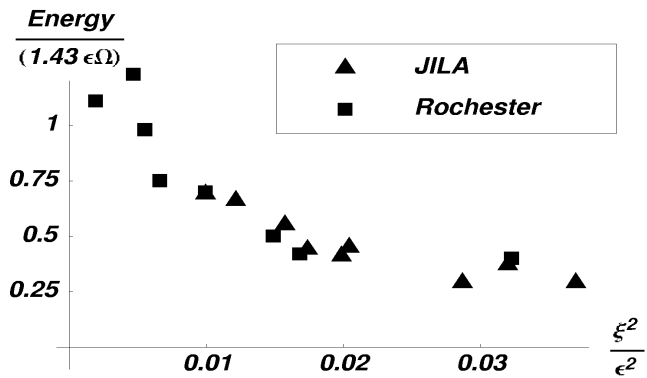}
\caption{Energy of the (1,0) mode vs. $\xi^2/\epsilon^2$. In Ref.~\cite{anglin} the
energy of this mode is given as 1.43$\epsilon\Omega$. }
\label{excitation_data}
\end{figure}
\begin{figure}
\includegraphics*[width=6cm,height=5cm]{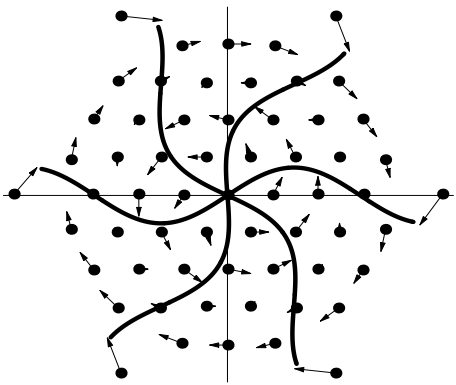}
\caption{Distortion of an array $\left(\xi^2/\epsilon^2=0.007\right)$ 
in a (1,0) mode. The dots represent the equilibrium positions of the vortices, and the 
straight arrows represent the actual calculated vortex displacements. The curve 
is a fit of a sine wave of wavelength 1.33$R_{TF}$ to the vortex displacements. }
\label{good_schematic}
\end{figure} 
\begin{figure}
\includegraphics*[width=6cm, height=5cm]{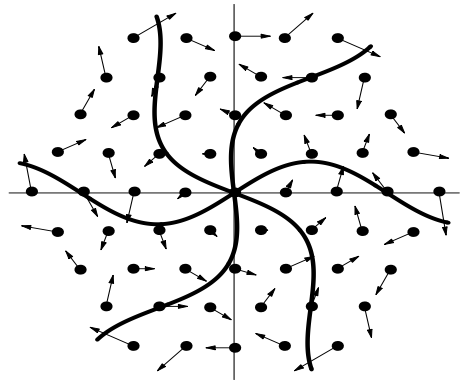}
\caption{Distortion of an array $\left(\xi^2/\epsilon^2=0.032\right)$ 
in a (1,0) mode. Compare with Fig.~\ref{good_schematic}, where 
$\left(\xi^2/\epsilon^2\right)$ is much
smaller. All symbols have the same meaning as in Fig.~\ref{good_schematic}.}
\label{bad_schematic}
\end{figure}
This technique however, has limitations. Even with a high-resolution movie, 
the excitations are usually difficult to tell apart by eye, especially as 
$\xi^2/\epsilon^2$ grows. The 
largely transverse Tkachenko waves become increasingly corrupted by sound waves traveling in 
the underlying fluid which scatter off the vortices. 
A more rigorous and reliable approach involves extracting 
the distortion $D_{n,m}$ from the fluctuation $\delta\psi$ 
induced by a particular excitation.

We therefore adopted a quantitative approach 
based upon determining the  instantaneous position of each vortex. A coarse distortion function $D_{n,m}$ can then be constructed by interpolating on the 
triangular lattice which specifies the groundstate. Along any circle centered at the origin, 
$r$ is constant
and the interpolated distortion $D_{n,m}$ defines a one dimensional 
function $d\left(\phi\right)$. The 
Fourier transform of $e^{-i\phi}d\left(\phi\right)$ 
then yields a peak near the 
value $m$ for the excitation labeled by $\left(n,m\right)$. This analysis
reveals one important result: the sense of precession of each vortex in an $m\neq0$ mode
may either be with or against the trap rotation. In these modes the sense of precession of a vortex
is a function of its radial coordinate only, and along any circle all vortices precess 
in the same sense.  This is a dramatic departure from the ideal situation of an array in a uniform, irrotational and 
incompressible fluid. In that case the vortices precess only against the trap rotation. We note that this has been analytically examined~\cite{jim}. 

Our central result is illustrated by Fig.~\ref{excitation_data}, where we make a quantitative 
comparision with the JILA experiment~\cite{jilat} and with recent analytic results~\cite{anglin}. 
We focus on the $(1,0)$ mode. 
As the parameter $\xi^2 / \epsilon^2$ is lowered (abscissa), we observe a quantitative deviation from analytic 
results (unity line on the ordinate) which is indistinguishable from that experimentally reported~\cite{jilat}. In proportion to this 
deviation, we find that the eccentricity of the elliptically polarized vortex oscillations in the Tkachenko wave is reduced and the individual vortex motion become almost circular. 
We illustrate this trend in Figs.~\ref{good_schematic} and~\ref{bad_schematic} 
(to be compared with Fig.~3 (b) in Ref.~\cite{anglin}). 
Notice that the vortex displacement vectors in Fig.~\ref{bad_schematic} have a more significant longitudinal 
projection than in Fig.~\ref{good_schematic} in which the ratio $\xi^2 / \epsilon^2$ is almost 
an order of magnitude larger.
In the opposite limit as $\xi^2 / \epsilon^2 \rightarrow 0$, we obtain very good
agreement with the analytical results.  We point out that it is not surprising that our energies begin to exceed the analytic
value (unity line). In the analytic treatment~\cite{anglin}, the energy of this 
mode was calculated to first order in $\epsilon$. 
For small $\xi^2 / \epsilon^2$, we consider
arrays for which $\epsilon\approx 0.2$ and for which higher order terms in $\epsilon$ could be important. 

In the limit of an incompressible fluid, the potential energy of the system is quenched and 
 Tkachenko waves emerge as the only nontrivial class of excitations of a 
vortex array. However, they are just one of several types of excitations which may 
occur within the compressible vortex arrays realized in experiments~\cite{sonin}. 
Of particular are the bulk modes~\cite
{svidfetter, choi, cozzini}, some of which can be very similar in appearance to Tkachenko waves. 
\begin{figure}
\includegraphics[width=8.0cm,height=8.0cm]{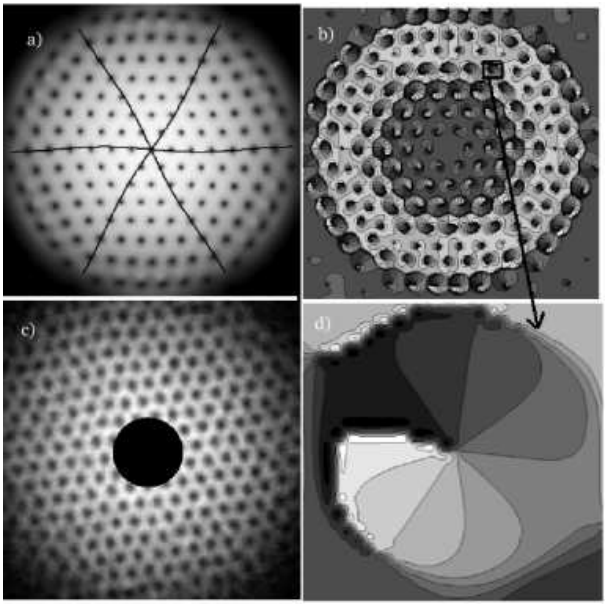}
\caption{a) A bulk (2,0) mode in action. The curve is a sine wave of wavelength 
1.33$R_{TF}$ predicted for the (1,0) Tkachenko wave. b) Phase plot of $\delta \rho$ 
Eq.~(\ref{density}). Notice the $\pi$ phase change as you radially cross a node. 
c) Approximate region in the array from which atoms are removed to excite 
the (2,0) bulk or (1,0) Tkachenko modes in the JILA experiment. 
d) Close-up of b) on the location of a vortex. 
In both of the phase plots above, phase changes from 0 to $2\pi$ as you go 
from dark to light regions.}
\label{bulk_motion}
\end{figure}
In Ref.~\cite{jilat}, the observation of a rapid mode which is visually 
indistinguishable from the Tkachenko (1,0) mode is reported.  We also idenitify such a mode 
at exactly the experimentally reported value of $2.25\omega_{\rho}$ - this is just the
second order bulk breathing or $(2,0)$ mode viewed in a rotating frame. 
During each half of the
linear oscillatory motion imposed on the vortices by the breathing action, 
the transverse (Magnus) force acts in opposite directions. On a microscopic level, 
elliptical motion of the vortices is observed. 
Macroscopically, the array twists around the center of rotation like a torsional 
pendulum as the density breaths radially.  We depict this second order 
`breathing' mode in Fig.~\ref{bulk_motion} a). Surprisingly, the sine wave radial 
distortion predicted~\cite{anglin} for the $(1,0)$ Tkachenko
mode is also in very good agreement with this $(2,0)$ bulk mode. 
To further illustrate this point, we make use of the complex density fluctuation 
defined for the $n$th excitation by~\cite{svidfetter}, 
\begin{equation}
\delta \rho= (\psi^* u_{n} - v_{n}^*\psi).  
\label{density}
\end{equation}
A casual comparision of the phase of $\delta \rho$ for this mode in Fig.~\ref{bulk_motion} b) 
with a schematic of the experimental probe in Fig.~\ref{bulk_motion} c) confirms the correct 
identification
of the mode excited in the experiment. A close up of this phase plot around the 
position of a vortex is shown in Fig.~\ref{bulk_motion} d). This describes a clockwise 
elliptical oscillation of the density fluctuation which has a local peak at the vortex core. 
A similar motion of the vortex at this site is implicated. We predict the existence of 
third and fourth order breathing modes at $2.55\omega_{\rho}$ 
and $3.30\omega_{\rho}$ respectively. This is a clear example of how the presense of 
a transverse force may blur the visual distinction of the largely longitudinal
modes from the transverse Tkachenko waves~\cite{sonin}. We have observed promising candidates
for the differential longitudinal waves which have been suggested previously~\cite{jilat}. 
In these relatively high energy excitations, vortex oscillations drive and 
are driven by large scale fluid density fluctuations. We shall report on these modes elsewhere. 
 
We thank Michael Banks for valuable computing assistance and G. S. Krishnaswami for insightful discussions. We are also
grateful to E. A. Cornell for permission to present the JILA data.
L. O. Baks. is a Horton Fellow. This work is supported by NSF, ARO and ONR.

\end{document}